# Statistics of electron-multiplying charge-coupled devices

**Brian M. Sutin**[*]

Jet Propulsion Laboratory/California Institute of Technology, 4800 Oak Grove Dr, Pasadena, CA 91109

**Abstract**

EMCCDs are efficient imaging devices for low surface brightness UV astronomy from space. The large amplification allows photon counting, the detection of events versus non-events. This paper provides the statistics of the observation process, the photon-counting process, the amplification, process, and the compression. The expression for the signal-to-noise of photon counting is written in terms of the polygamma function. The optimal exposure time is a function of the clock-induced charge. The exact distribution of amplification process is a simple-to-compute powered matrix. The optimal cutoff for comparing to the read noise is close to a strong function of the read noise and a weak function of the electron-multiplying gain and photon rate. A formula gives the expected compression rate.

\*E-mail: brian.m.sutin@jpl.nasa.gov

## 1 Science Motivation

The Earth's atmosphere is essentially opaque to ultraviolet (UV) radiation at wavelengths shorter than about 300 nm. All sorts of interesting astrophysical phenomena exist to be imaged in the UV, such as emission lines from interstellar gas, redshifted H-α from galaxies, and zodiacal light caused by Rayleigh scatter of sunlight. Most of the UV sources mentioned above have very low UV surface brightness, so observation requires an imaging telescope above the Earth's atmosphere that integrates for long periods, coupled with an imaging detector with very high efficiency. Typical photon rates are such that observing for a second of integration time results in typical photon counts on the order of a photon per 1,000 pixels. As space-based missions are limited in the returned data, compression of data consisting of non-detections is relevant.

This paper assumes a mission similar to the Polarized Zodiacal Light Experiment (PoZoLE) concept [1], intended to observe our solar system's zodiacal light's UV polarization from Low-Earth Orbit (LEO). The zodiacal cloud is our solar system's biggest structure visible to the unaided eye, yet its constituent dust particles' origins are controversial, with a wide range of proposed divisions between sources in the asteroids and Jupiter Family comets. Furthermore, any contribution from Oort Cloud comets is poorly constrained. The zodiacal cloud gains new meaning with the Astro2020 Decadal Survey recommending that NASA develop a large space telescope to characterize potentially Earth-like planets around nearby stars. The large telescope will seek signs of oxygen in the atmospheres of exo-Earths by the Hartley absorption band of ozone in the near-ultraviolet — the same absorption that helps protect us on the Earth's surface from sunburn. However, the exo-Earths and their ozone may be obscured by the host stars' light scattered from their extrasolar analogs of our zodiacal cloud. As the only such cloud where the source bodies can be tracked, our own interplanetary dust is a key to learning how the system architecture governs the cloud's appearance. PoZoLE will observe our solar system's dust from the outside in the near-UV. Here, it will enable designing a large telescope to best pick out an Earth analog against a dust cloud like our own.

The key to turning observations from Earth orbit into a view of our zodiacal cloud from the outside is to measure the polarization of the sunlight scattered from the cloud's dust particles. PoZoLE will "observe" the modeled near-UV appearance of a proposed structure for the zodiacal cloud against a synthetic Milky Way background, and show the measurements are sufficient to determine the cloud's shape and size, and estimate how much of the UV-scattering dust comes from each source family: asteroids, Jupiter Family comets, and Oort Cloud comets, with their differing orbits.

## 2    What is an EMCCD?

Electron-Multiplying Charge-Coupled Devices (EMCCDs) are the enabling technology for space-based UV imaging, especially with enhanced UV quantum efficiency. The EMCCD is a CCD modified so that can achieve high signal-to-noise ratio (SNR) by rendering the read noise effectively zero. Compared to conventional CCDs, EMCCDs have an additional serial register (604 extra charge-coupled "pixels" (see Figure 1), where one of the register clocks is replaced by a high-voltage clock (25–50 V). The higher voltage causes a multiplication process that stochastically turns one electron into many, resulting in thousands of electrons at the output amplifier. This allows detection of single-photon events by thresholding above amplifier read noise [2].

In the amplification process, if the charge transfer from one register to the next is done with enough input energy, a signal electron will knock loose another electron. This is the Avalanche-Photodiode (APD) effect [3]. Repeating this process makes an "APD staircase" [3], magnifying the signal by having some extra CCD-like charge transfers explicitly run using overdrive. Since the digitization takes place after magnification, the read noise is proportionally smaller when compared to the original, unamplified signal.

EMCCDs are generally run in "photon counting" or PC mode, meaning that the detector is read out often enough that the expected signal counts in each pixel are $\ll 1$. After magnification, the final counts are generally either very large or very small, signifying detecting or not detecting a signal photon. This is similar to a photomultiplier tube. The downside, as compared to a PMT or ICCD, is that the dark current and Clock-Induced Charge (CIC, see §6) may also add significant noise. The EMCCD data output is then a combination of science signal, dark current, CIC, and readout noise.

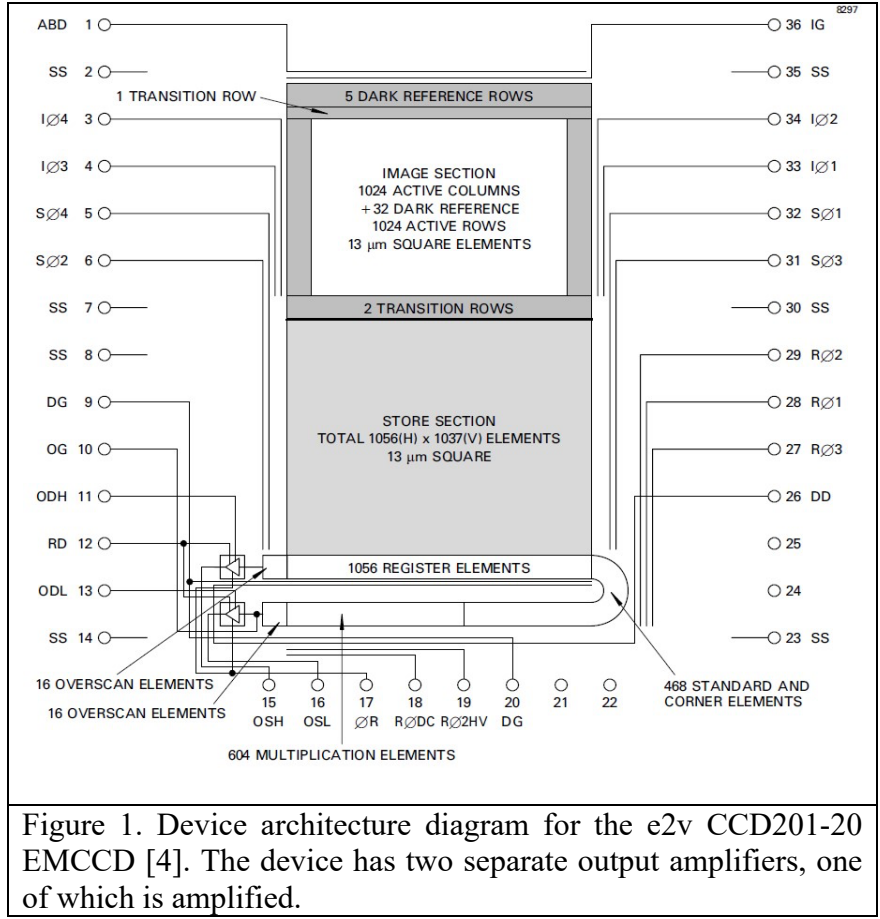

Figure 1. Device architecture diagram for the e2v CCD201-20 EMCCD [4]. The device has two separate output amplifiers, one of which is amplified.

The UV detection efficiency of these detectors is extremely enhanced by delta-doping, a surface treatment that increases quantum efficiency, and metal-dielectric filters directly applied to the detector surface. Through years of collaborative effort, JPL and Teledyne e2v have demonstrated and perfected the process flow for developing high-efficiency UV silicon detectors through several programs [5], including Faint Intergalactic Red-shifted Emission Balloon (FIREBall-2) [6-8], SPARCS [9-11], and NASA-SAT and JPL-funded efforts. Roman CGI has extensively tested and qualified the CCD201-20 [12]. PoZoLE's UV metal-dielectric filter is identical to one of the SPARCS filters, and is optimized for a 250-300 nm passband.

## 3 Probability Notation

The rest of this paper mainly consists of probability theory related to processing and understanding the signal from these detectors. The notation used is

$X$, $X_Y$ - a random variable; i.e., a variable that can take on randomly generated values
$P(X)$ - a probability distribution
$E(X)$ - the expected value of $X$, also known as the first moment or mean
$V(X)$ - the variance of $X$, equal to the second moment of $X$ minus $E(X)^2$

As the intended target of this paper is engineers and scientists rather than mathematicians, neither mathematical rigor nor mathematically rigorous notation are prioritized.

## 4    Poisson Distributions

For independent events arriving at a detector, the Poisson distribution is appropriate to use. For a Poisson distributed random variable $X$, $E(X) = V(X) = \lambda$, and $X$ is distributed as

$$P(X = k) = \frac{\lambda^k e^{-\lambda}}{k!} \tag{1}$$

Similar to the Gaussian assumption, the SNR goes as the square root of the counts as a result of $E(X) = V(X)$. For no events

$$P(X = 0) = e^{-\lambda}$$
$$P(X > 0) = 1 - e^{-\lambda} \tag{2}$$

Useful is the Cumulative Distribution Function (CDF), which is

$$F_X(n) = P(X \leq n) = \sum_{k=0}^{n} \frac{\lambda^k e^{-\lambda}}{k!} = Q(n+1, \lambda) \tag{3}$$

Here $Q(n, \lambda)$ is the Regularized Gamma Function; e.g., evaluated in Excel as

$$Q(n, \lambda) = 1 - GAMMA.DIST(\lambda, n, 1, TRUE) \tag{4}$$

## 5    Photon Counting

Consider an ideal photon-counting detector with infinite gain. If any photons at all are detected, then the detector returns 1, otherwise 0. Let the random variable $X_W$ be the number of electrons in the well of a pixel. Using Equation (2), the expected value (and second and all higher moments) of this process is then

$$\begin{aligned} E(Ideal\ PC) &= \sum_{w=1}^{\infty} (1)\, P(X_W = w) \\ &= P(X_W > 0) \\ &= 1 - e^{-\lambda} \end{aligned} \tag{5}$$

In practice, a single exposure is not useful. Rather, $\lambda$ is estimated by taking a sequence of $N$ exposures, each in PC mode. Define the $\varepsilon$ as the probability of detection,

$$\varepsilon = 1 - e^{-\lambda} \tag{6}$$

The number of exposures with a signal detection are then given by the binomial distribution

$$P(PC_N = c | \varepsilon) = \binom{N}{c} \varepsilon^c (1 - \varepsilon)^{N-c} \tag{7}$$

Here $c$ is the total number of counts for the $N$ exposures. If $\varepsilon$ is ½, then this is the distribution of the number of heads after $N$ flips of a fair coin. What we want to know is $\varepsilon$, which is on the wrong side of the equation. Using Bayes' theorem

$$P(\varepsilon|PC_N = c) = \frac{P(PC_N = c \,|\varepsilon)P(\varepsilon)}{P(PC_N = c)} \tag{8}$$

We need a prior for $P(\varepsilon)$ and select the most commonly used prior for the binomial distribution, the Beta distribution, given by

$$P(\varepsilon) = \frac{\varepsilon^{\alpha-1}(1-\varepsilon)^{\beta-1}}{B(\alpha,\beta)} \tag{9}$$

The denominator is the Beta function $B(\alpha,\beta) = \Gamma(\alpha)\Gamma(\beta)/\Gamma(\alpha+\beta)$, and the distribution is valid over $\{\alpha > 0, \beta > 0\}$. A common choice is the uniform prior $\{\alpha = 1, \beta = 1\}$. For a reader unfamiliar with Bayesian statistics, the prior can be used to bias the result using prior knowledge (thus the name). However, the actual choice of prior makes little difference with sufficient data. More useful is to reverse this last statement; if the result of an observation depends significantly on the choice of prior, then the data quantity is insufficiently robust. Using the above prior,

$$P(\varepsilon|PC_N = c) = \frac{\varepsilon^{c+\alpha-1}(1-\varepsilon)^{N-c+\beta-1}}{B(c+\alpha, N-c+\beta)} \tag{10}$$

We did not need to explicitly calculate the Bayesian denominator $P(PC_N = c)$. The denominator is determined by noticing that the binomial distribution combined with a Beta distribution is another Beta distribution, so the new denominator is the normalization factor from the new Beta distribution.

The expression for the moments of $\lambda$ is found by using the probability distribution for $\varepsilon$ to find the powers of $\lambda$ written as a function of $\varepsilon$.

$$M_n(\lambda) = \int_{\varepsilon=0}^{1} (-\ln(1-\varepsilon))^n \frac{\varepsilon^{c+\alpha-1}(1-\varepsilon)^{N-c+\beta-1}}{B(c+\alpha, N-c+\beta)} \, d\varepsilon \tag{11}$$

A change of variable gives, to within a sign, the formula for the geometric moments of the Beta distribution,

$$M_n(\lambda) = \int_{x=0}^{1} (-\ln x)^n \frac{x^{N-c+\beta-1}(1-x)^{c+\alpha-1}}{B(N-c+\beta, c+\alpha)} \, dx \tag{12}$$

So, the mean and variance of $\lambda$ are (within a sign) the geometric mean and geometric variance of a Beta distribution. The expected value can be written in terms of the digamma function $\psi$,

$$E(\lambda) = -M_1(\lambda) = \psi(N+\alpha+\beta) - \psi(N-c+\beta) \tag{13}$$

and the variance in terms of the trigamma function $\psi_1$,

$$V(\lambda) = M_2(\lambda) - M_2^2(\lambda) = \psi_1(N-c+\beta) - \psi_1(N+\alpha+\beta) \tag{14}$$

The polygama function $\psi_n$ may seem scary but is just another transcendental function available in language math libraries or evaluated with a series approximation.

## 6    Optimal Exposure Time

$\lambda$, the expected number of counts in an exposure, depends on the exposure time, while the relevant quantities for an observation are the rate of photons per unit time and the total exposure time available. The obvious choice is to have a very large number of exposures and make the exposure time arbitrarily small. However, in practice, An EMCCD has a noise source called Clock-Induced Charge (CIC) that is added to every exposure [8,13], where an extra charge is added during clocking of the charge-coupled pixels. CIC has been measured as low as $7\times10^{-4}$ e-/pixel/readout [14], while achieving is $5\times10^{-3}$ e-/pixel/readout is relatively easy.

CIC is not constant across the detector. For a 1k×1k-pixel frame transfer device, the number of pixel-to-pixel transfers can vary from 1k to 3k (frame transfer plus row number plus column number) before arriving at the readout and amplification chain (see Figure 1).

For this section, define the following variables:
- $T$- the total exposure time in seconds
- $\lambda_{CIC}$- the CIC count rate in counts per exposure
- $\eta$- the photon count rate from the science signal source in counts per second

The variable $\eta$ is the fundamental astronomical parameter being measured. Our expected counts per exposure $\lambda$ is then

$$\lambda = \lambda_{CIC} + \eta T/N \tag{15}$$

The expected photon-counting event detection rate is

$$\bar{\varepsilon} = 1 - e^{-\lambda_{CIC} - \eta T/N} \tag{16}$$

In order to avoid the unpleasant polygama function, we will find the optimal number of exposures $N$ by maximize the SNR with respect to $\varepsilon$ rather than $\lambda$. This should be acceptable, as CIC is only critical for small $\lambda$. Choose the number of exposures $N$ to be $N \gg \alpha$ and $N \gg \beta$ (or choose $\alpha = \beta = 1$). Since $\varepsilon$ is Beta distributed, we know the mean and variance, and thus the SNR,

$$\begin{aligned} E(\bar{\varepsilon}) &= c/N \\ V(\bar{\varepsilon}) &= \frac{(c/N)(1 - c/N)}{N} \\ SNR(\bar{\varepsilon}) &= \left(\frac{c}{1 - c/N}\right)^{\frac{1}{2}} \end{aligned} \tag{17}$$

The SNR evaluated at the expected $\bar{\varepsilon}$ is then

$$SNR(\bar{\varepsilon}) = N^{\frac{1}{2}}\left(e^{\lambda_{CIC} + \eta T/N} - 1\right)^{\frac{1}{2}} \tag{18}$$

Setting the derivative with respect to $N$ to zero gives the transcendental equation

$$(1 + \lambda_{CIC} - \lambda)e^{\lambda_{opt}} = 1 \qquad (19)$$

Over the typical region of $10^{-3} < \lambda_{CIC} < 5 \times 10^{-3}$, $\lambda_{opt}$ is well approximated by the fit

$$\lambda_{opt} \approx \frac{3}{2}\lambda_{CIC}^{\frac{1}{2}} \qquad (20)$$

So, for CIC ~ 0.001, the optimal number of well counts per exposure is $\lambda_{opt}$ ~ 0.045 counts for best SNR. Given that CIC for current state of the art EMCCD controllers is in the range of 0.001 to 0.005, lambda for optimal SNR is restricted to (0.045 to 0.10). Note that other considerations might lead to using shorter exposure times than for optimal SNR. For example, bright stars in the field can cause "blooming", where the wells of the amplification staircase elements overfill and spill charge into adjacent elements. The exposure time may also be limited by the spacecraft stability for applications where spatial resolution is important.

## 7    Comparison to Standard Mode

For bright sources, the detector frame rate may not be high enough to achieve the optimal count rate lambda per exposure. In this case, observing with standard mode might provide better SNR. The SNR in standard mode is

$$SNR(SM) = \frac{\eta T}{(\eta T + \hat{\mu})^{\frac{1}{2}}} \qquad (21)$$

Here $\hat{\mu}$ is the readout noise for standard mode, which may be different than the readout noise $\mu$ for PC mode if the EMCCD device has separate readout amplifiers (see Figure 1). Given the total number of expected events $\eta T$ for an observation and a known $\lambda_{CIC}$, the number of PC-mode exposures is

$$N = \eta T / (\lambda - \lambda_{CIC}) \qquad (22)$$

The expected number of PC counts is

$$E(c) = N - Ne^{-\lambda} \qquad (23)$$

Assuming $N$ is large enough that the prior is irrelevant and replacing $c$ with the expected counts, the PC-mode SNR is

$$SNR(PC) = \frac{\psi(N) - \psi(Ne^{-\lambda})}{\left(\psi_1(Ne^{-\lambda}) - \psi_1(N)\right)^{\frac{1}{2}}} \qquad (24)$$

Equations (21) and (24) can be compared to choose which mode is best. A worked example is shown in Figure 2. In practice, the standard-mode exposure time $T$ is limited by cosmic rays, while for PC mode the exposure time $T/N$ is limited by readout speed.

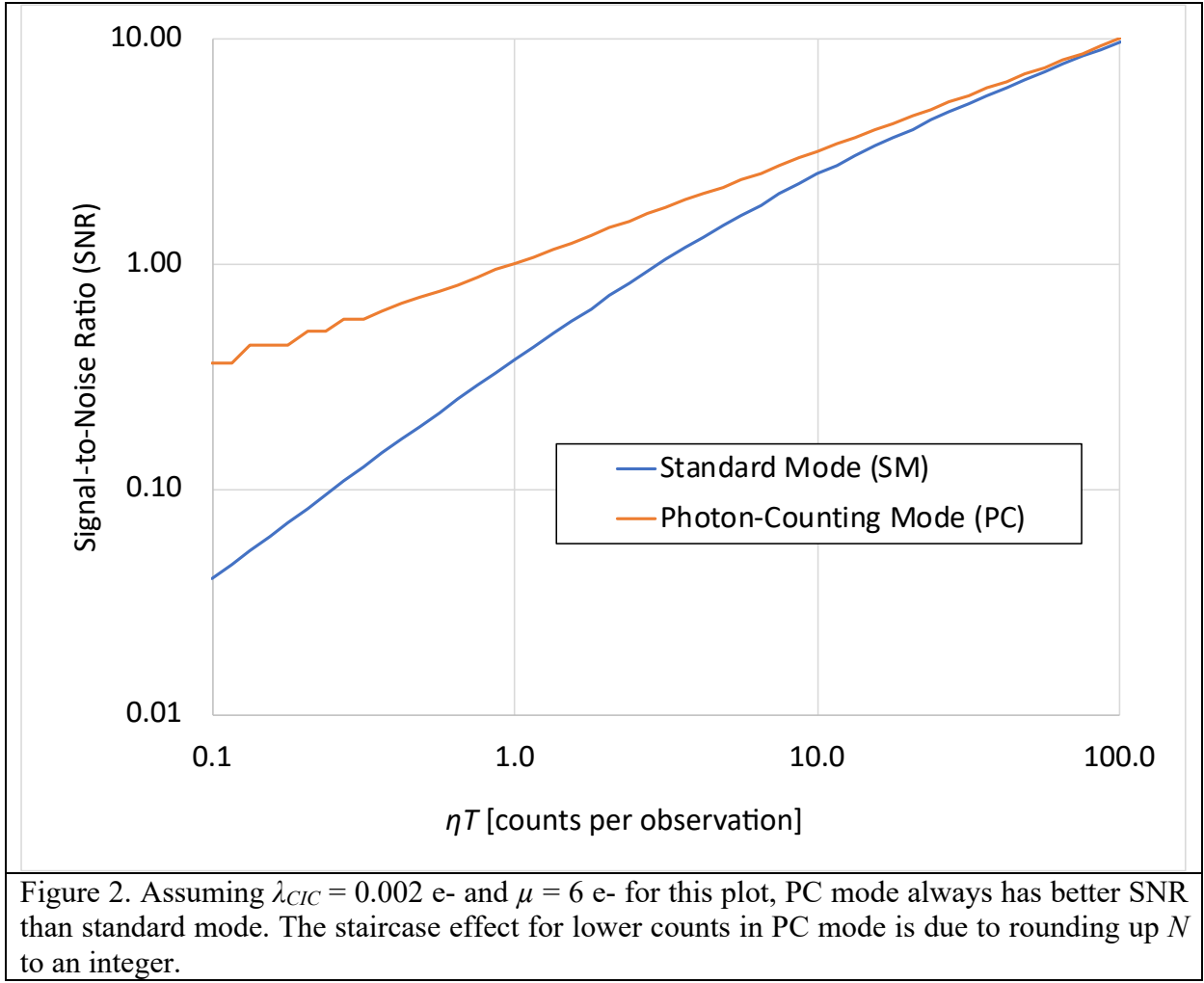

Figure 2. Assuming $\lambda_{CIC} = 0.002$ e- and $\mu = 6$ e- for this plot, PC mode always has better SNR than standard mode. The staircase effect for lower counts in PC mode is due to rounding up $N$ to an integer.

## 8 Amplification

The amplification process of EMCCD amplification is similar to the amplification process in avalanche photodiodes (APD) [3,15-19]. The amplification of a single electron from *Matsuo* 1984 [15] (hereafter 'Matsuo') is given by a recurrence relation for each stage $m$ by

$$
\begin{aligned}
P_m(n) &= (1-Q)P_{m-1}(n) + Q\sum_{k=0}^{n} P_{m-1}(n-k)P_{m-1}(k) \\
P_0(n) &= \delta_{1,n}
\end{aligned}
\tag{25}
$$

The distribution of counts for each stage $P_m(n)$, where $n$ is the number of counts, depends on the distribution of counts at the previous stage. What happens if the initial distribution of counts is not restricted to a single count, or $P_0(n) \neq \delta_{1,n}$? Then the recurrence relation spectacularly fails. However, the most important issue with this recurrence relation is that it physically makes no sense; the relation connects the opposite tails of the distribution to generate the next stage. So, although the relation may hold true, it is more of a mathematical curiosity than anything else.

The APD staircase effect can be described as a process where, at each step, there is a chance when transferring the charge that an electron generates a second electron. Now make two assumptions. First, the electrons are not aware of each other, equivalent to the process at each step being linear in the number of electrons generated on average. Second, assume that an original electron can never generate two or more additional electrons, equivalent to assuming that the energy of an electron, once spent on generating a second electron, is gone. With these two assumptions, we can write down a recurrence relation.

Consider at some step $m$ in the APD staircase, the distribution of electrons $P_m(n)$. How was this distribution generated if the previous step had $k$ electrons? Clearly, $n - k$ electrons each generated a single electron (first assumption), while $k - (n - k) = 2k - n$ electrons did not. Let $Q$ be the probability of an electron generating a second. Since the electrons are independent (second assumption), the applicable distribution is the binomial distribution.

$$P_m(n) = \sum_{k=\lceil n/2 \rceil}^{n} \binom{k}{n-k} (1-Q)^{2k-n} Q^{n-k} \, P_{m-1}(k) \tag{26}$$

This expression looks unwieldly; however, the right-hand side is nothing more than a fixed matrix multiplication to get from one stage of the APD to the next. The matrix is

$$B_{nk} = \binom{k}{n-k} (1-Q)^{2k-n} Q^{n-k} \qquad k \in [\lceil n/2 \rceil, n] \tag{27}$$

Once the matrix $B_{nk}$ is formed with a size corresponding to the largest electron counts of interest, the recurrence relation becomes

$$P_m(n) = B_{nk} \, P_{m-1}(k) \tag{28}$$

The final amplified distribution is then

$$P_m(n) = (B_{nk})^m \, P_0(k) \tag{29}$$

Computing $P_m(n)$ is only a few lines of MATLAB code,

```
% Create the transform matrix for a single stage of a staircase APD
Q = gain^(1/stages) - 1;
for n = (0:nmax);
  for k = (0:nmax);
    if( (k <= n) && (n-k <= k) )
      B(k+1,n+1) = exp( gammaln(k+1) - gammaln(n-k+1) - gammaln(2*k-n+1)
+ (2*k-n)*log(1-Q) + (n-k)*log(Q) );
    end % k
  end % n
  % Transform initial distribution to the final distribution
  P = Po*BB^stages; % Po and P are row vectors here
```

This recurrence relation gives identical results to Matsuo for an initial distribution of a single electron $P_0(n) = \delta_{1,n}$, while also, unlike Matsuo, working for any initial distribution.

## 9    Comparison to Basden

*Basden* 2003 [20] (hereafter 'Basden') gives the formula for the distribution of final amplified counts $X_A$ as (Basden, Equation 1)

$$\tilde{P}(X_A = a \mid X_W = w) = \frac{a^{w-1} e^{-a/G}}{G^w (w-1)!} \qquad a \geq w, w > 0$$

$$\tilde{P}(X_A = 0 \mid X_W = 0) = 1$$

(30)

$G$ here is the gain of the entire APD staircase, so the gain at stage $m$ would be $G_m = (1+Q)^m$.

Basden's derivation is based on assuming the first assumption above (independent electrons) and a few approximations. We write $\tilde{P}$ because the formula is not a probability distribution; the sum over all values is not equal to unity. The formula has the functional form of an Erlang distribution, but the Erlang distribution is a <u>continuous</u> distribution in $a$. The Erlang distribution is the distribution of the sum of $w$ independent identically distributed (IID) exponentially distributed random variables. This is a hint that the Basden equation could be derived as a sum of IID geometrically distributed random variables. In fact, a plot of the Matsuo distribution for a single initial electron does look very similar to an exponential distribution.

The recurrence relation of Equation (25) deviates from that of Equation (26), mostly for larger counts (Figure 3). Basden explicitly states they their approximation should not be used in this region.

In the limit as the number of stages $m$ becomes large and the amplification $Q$ becomes small for fixed $G$, a closed-form solution in terms of $G$ based on Equation (26) should exist. At a minimum, the recurrent relation can be written in terms of the eigenvectors and eigenvalues,

$$B_{nk}^m = V \Lambda^m V^{-1}$$

(31)

Since the matrix $B_{nk}$ is triangular, the diagonal elements are the eigenvalues. Taking the limit as $m$ goes to infinity for fixed $G$ gives the eigenvalues as

$$\Lambda_i = G^{-i}$$

(32)

In the limit, the eigenvectors $V$ are independent of $G$.

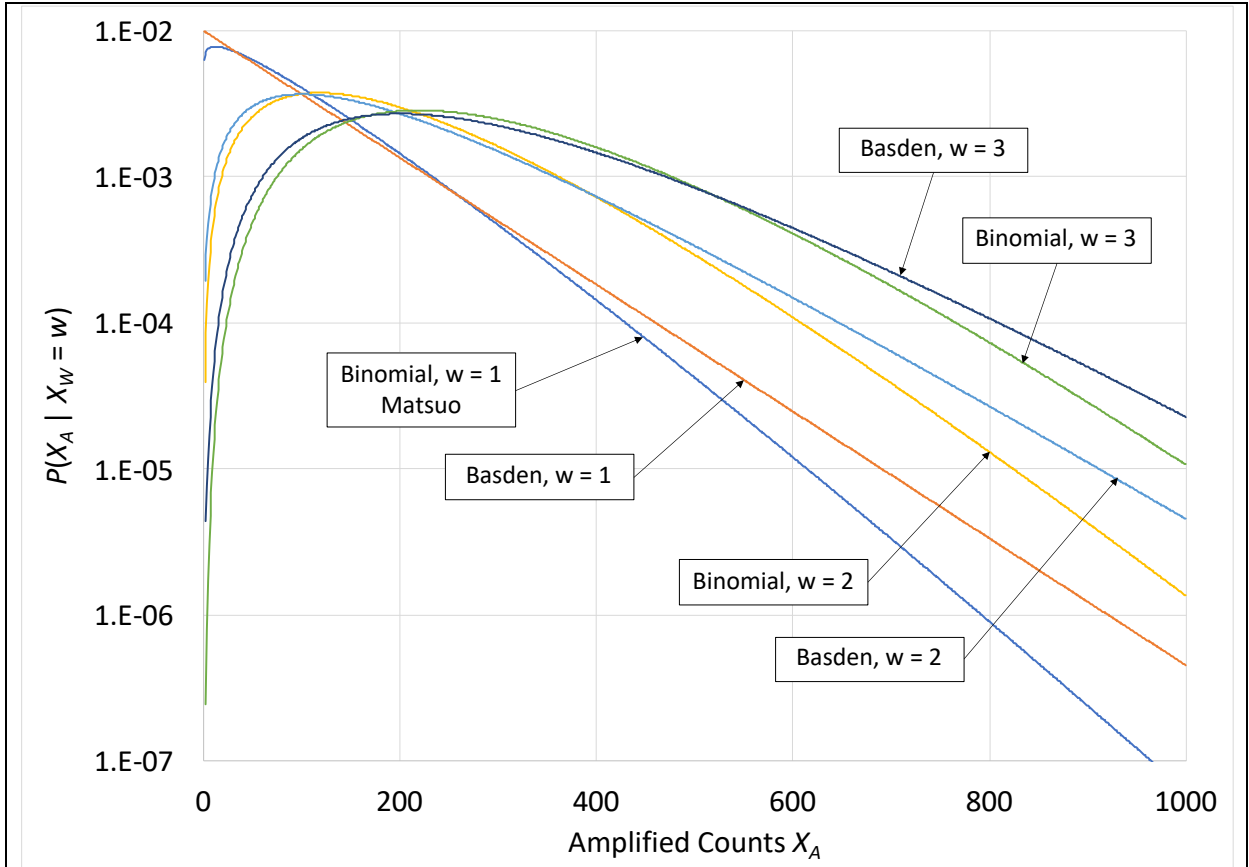

Figure 3. The Basden approximation is close to the binomial model for moderate counts but fits poorly for the less likely counts. For this plot, the gain is 100, with 50 stages. 'Matsuo' is Equation (25), 'Basden' is Equation (30), and 'Binomial' is Equation (26).

## 10 Combining Signal with Read Noise

In practice, we add the amplified well counts to the read noise $X_R$. Read noise is also Poisson distributed, so

$$P(X_R = y) = \frac{\mu^y e^{-\mu}}{y!} \tag{33}$$

The expected value of the read noise $\mu$ is trivially measured by taking a zero-exposure-time frame or alternatively running the ADC without advancing the CCD charge. Since the read noise and amplified counts are independent, the distribution of the sum read out by the detector is

$$P(X_A + X_R = c) = \sum_{a=0}^{c} P(X_A = a) P(X_R = c - a) \tag{34}$$

Inserting in the expressions from Equations (1), (29), and (33),

$$P(X_A + X_R = c) = \sum_{a=0}^{c} \frac{\mu^{c-a} e^{-\mu}}{(c-a)!} B_{ak}^m \frac{\lambda^k e^{-\lambda}}{k!} \tag{35}$$

## 11 PC Cutoff Setting

The algorithm used for photon counting (PC) is that $c$ is considered a single count if higher than some threshold $\tau$. The expected value (and second and all higher moments) of this process is then

$$E_\tau(PC) = \sum_{c=\tau+1}^{\infty} (1) \, P(X_A + X_R = c) \tag{36}$$

Since the argument of the summation is a probability distribution over $c$, this is equal to

$$E_\tau(PC) = 1 - \sum_{c=0}^{\tau} P(X_A + X_R = c) \tag{37}$$

Substituting in Equation (35) and swapping the summations gives

$$E_\tau(PC) = 1 - e^{-\lambda} \sum_{a=1}^{\tau} \sum_{m=1}^{\tau-a} \frac{\mu^m e^{-\mu}}{m!} B_{ak}^m \frac{\lambda^k}{k!} \tag{38}$$

The sum over $m$ is again the Regularized Gamma Function from Equation (3),

$$E_\tau(PC) = 1 - e^{-\lambda} \sum_{a=0}^{\tau} Q(\tau - a + 1, \mu) B_{ak}^m \frac{\lambda^k}{k!} \tag{39}$$

The optimal choice for the cutoff parameter $\tau$ is such that the read noise does not bias the PC process. This is equivalent to setting $\tau$ such that false positives equal false negatives. With this choice, lost counts from low-amplification signal counts balances added counts from high noise. The photon counting method is then an "ideal" photon counter, as in §5. One can prove that setting false positives to false negatives is equivalent to the read noise having no effect, and so the expected value of $\varepsilon$ with noise is equal to $\varepsilon$ without noise, or

$$E_\tau(PC) \cong \varepsilon \tag{40}$$

Let $\hat{\tau}$ be the optimal $\tau$ that makes Equation (39) satisfied. Then

$$\sum_{a=0}^{\hat{\tau}} Q(\hat{\tau} - a + 1, \mu) B_{ak}^m \frac{\lambda^k}{k!} = 1 \tag{41}$$

Over the range of typical values for $\lambda$, $\mu$, and $G$, the optimal $\hat{\tau}$ is well fit to within 5% by

$$\hat{\tau} \approx 1.72 \mu^{0.88} (G/\lambda)^{0.03} \tag{42}$$

$\hat{\tau}$ only has a weak dependence on $\lambda$ and $G$. The residuals of the fit are due to the $\mu$ dependence not being sufficiently well-fit with a power law. The residuals are shown in Figure 4.

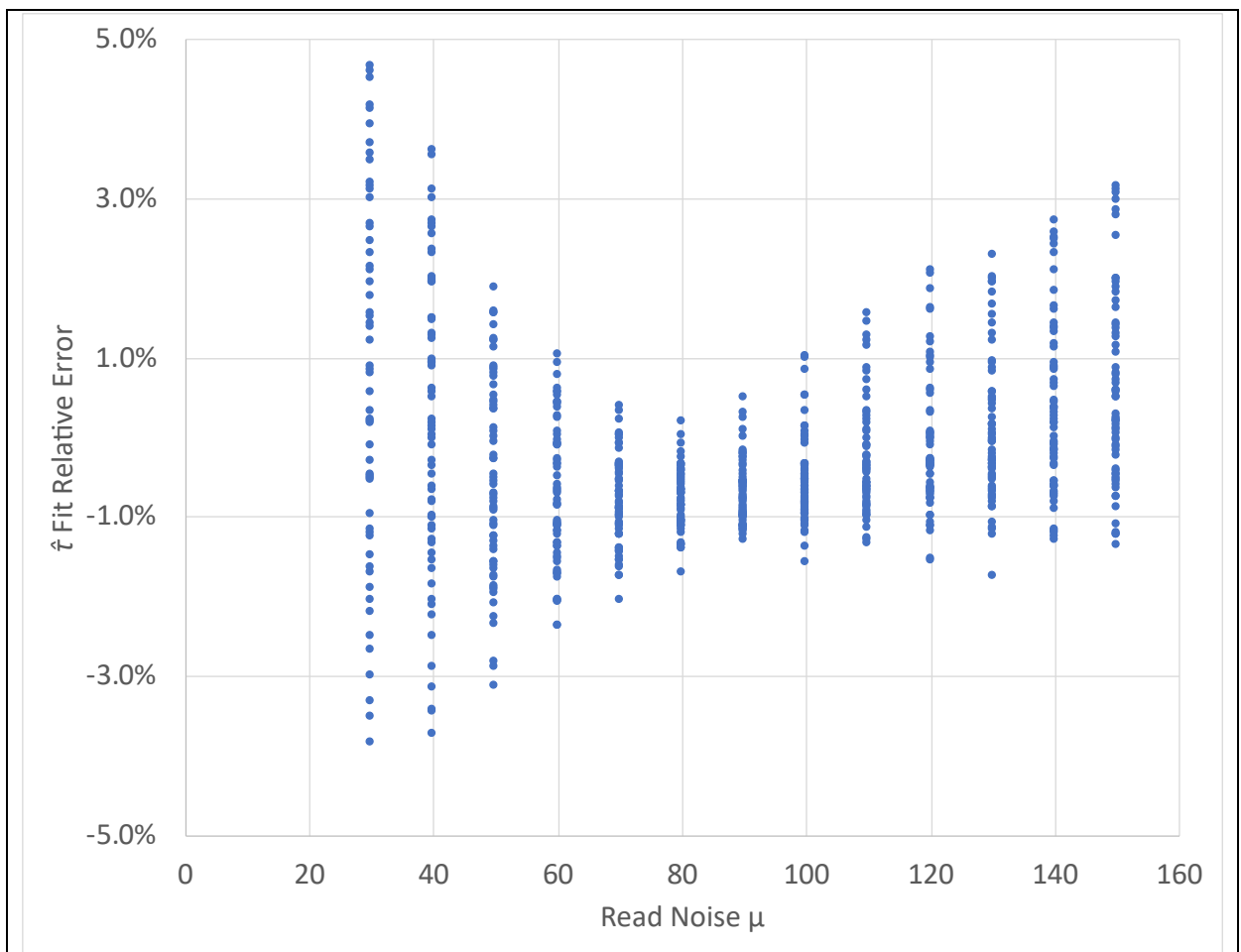

Figure 4. The residuals of the Equation (42) fit to the cutoff $\tau$ in Equation (41) are within 2%. The scatter is due to variations in $\lambda$ over 0.04 to 0.1 and $G$ over 10 to 1000.

## 12 Compression of Photon-Counting Mode Data

For a typical photon-counting exposure, the expected image data is a list of binary values, the vast majority of which are zeros. Since CIC may be a significant contributor to the signal and is unstructured random noise, compression algorithms based on entropy (structure in the image) are likely not optimal. On the other hand, an algorithm which depends on most of the data being zero might be.

If the events in each pixel are independent, then the statistics are Poisson distributed with a "rate" or expected number of pixels between events, of $\lambda$. For a detector where the noise is CIC-dominated and CIC is approximately 0.001 e-/pix/frame, the number of pixels between each photon detection is about 1/(0.001) = 1000.

The simplest compression method for sparse binary data is to count zeros between events. This nominally gives a compression ratio of approximately the number of pixels divided by the number of bits needed to encode the number, or $\sim(1/\lambda)/\log_2(1/\lambda)$. A compression ratio roughly on the order of 100:1 should be possible for the above CIC alone.

This algorithm is a variation of Run-Length Encoding (RLE), which is to count the number of bits before a change. The heritage of the RLE algorithm stretches back to at least the early 1960's [21]. For mostly empty data, the appropriate standard compression algorithm is the Run-Length Limited (RLL) variation, which has been common for encoding hard disks and optical storage since the 1980's, and only counts zeros. We use (0,$b$) RLL.

The encoding algorithm for RLL(0,$b$) is as follows:
1. Count number of zeros until a non-zero.
2. For an b-bit encoding, don't count more than $2^b$-2 zeros.
3. If no non-zero encountered, then return $2^b$-1.
4. Else, return number of zeros.

For the data generated by the photon-counting process, we can compute the expected compression ratio. Equation (1) gives the probability for an event (a one) and a non-event (a zero) for an image pixel. Assuming independent events, the probability of $n$ zeros followed by a non-zero is

$$P(X_{\leq n} = 0, X_{n+1} = 1) = e^{-n\lambda}(1 - e^{-\lambda}) \tag{43}$$

The expected number of inputs bits consumed by a $b$-bit encoding is then the probability of all zeros (consuming one encoding value), or some run of zeros followed by a non-zero.

$$E(\text{bits consumed}) = 2^b P(X_{\leq 2^b} = 0) + \sum_{n=0}^{2^b-1} (n+1)P(X_{\leq n} = 0, X_{n+1} = 1) \tag{44}$$

Simplifying,

$$E(\text{bits consumed}) = 2^b e^{-2^b\lambda} + (1 - e^{-\lambda}) \sum_{n=0}^{2^b-1} (n+1)e^{-n\lambda} \tag{45}$$

Using the derivative of the formula for the sum of a geometric series, this simplifies to

$$E(\text{bits consumed}) = 1 + 2^b e^{-2^b\lambda} - 2^b e^{-(2^b-1)\lambda} + \frac{e^{-\lambda} - e^{-2^b\lambda}}{(1 - e^{-\lambda})} \tag{46}$$

The final compression ratio is then $E$(bits consumed)/$b$.

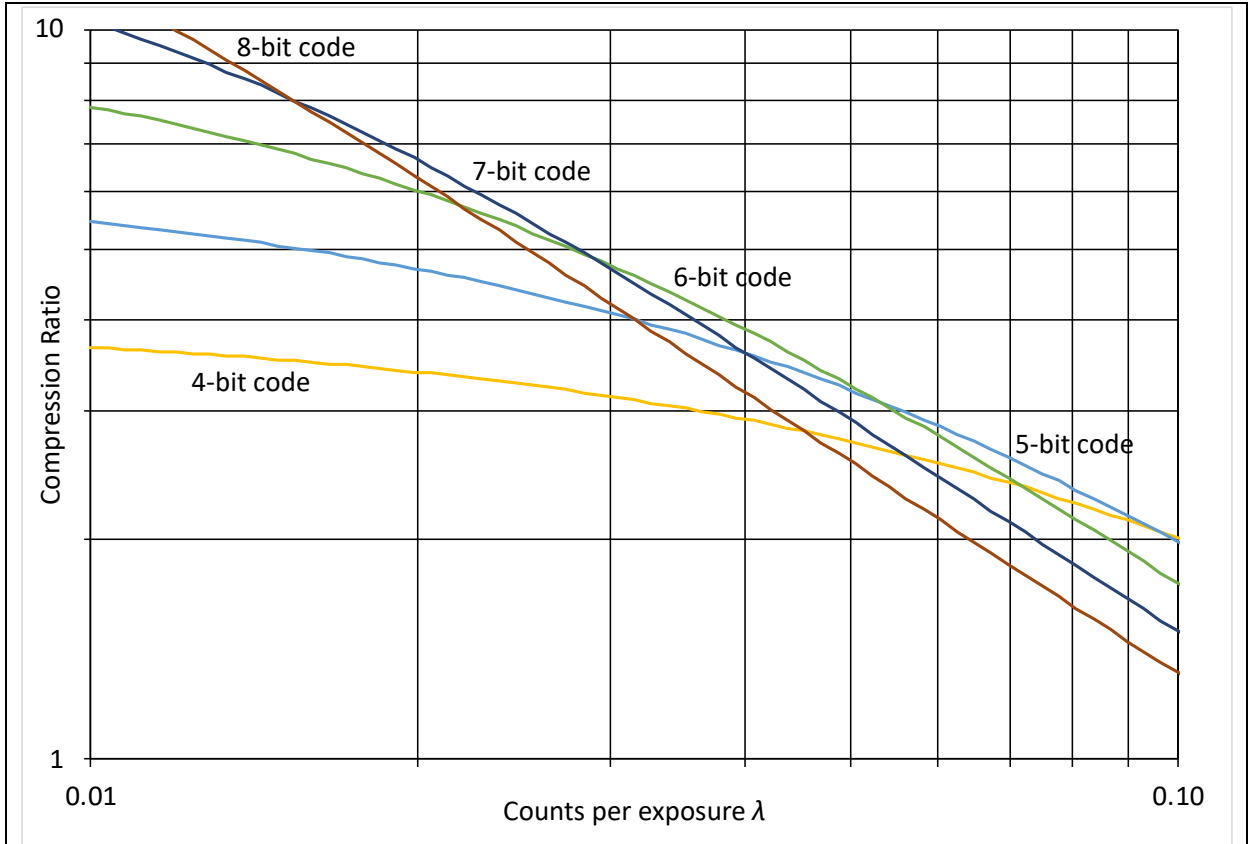

Figure 5. An encoding with 5-bit codes has the highest compression ratio for $0.053 < \lambda < 0.095$, which encompasses most of the range of $\lambda$ for optimal SNR from §6. The most likely lossless compression assuming optimal SNR is thus from 2:1 to 3:1. The optimal compression ratio for $\lambda < 0.01$ is well-fit by the approximation $0.264 \times \lambda^{-0.825}$.

The boundaries in $\lambda$ for the ranges of optimal compression are found by looking for the $\lambda$ where compression$(\lambda,b)$ = compression$(\lambda,b+1)$. The results are shown in Table 1.

Table 1. Optimal code length for given counts per exposure $\lambda$.

| Code Length | $\lambda_{lo}$ | $\lambda_{hi}$ |
|---|---|---|
| 2 | 0.2272 | 0.4114 |
| 3 | 0.1636 | 0.2272 |
| 4 | 0.0955 | 0.1636 |
| 5 | 0.0531 | 0.0955 |
| 6 | 0.0288 | 0.0531 |
| 7 | 0.0154 | 0.0288 |
| 8 | 0.0082 | 0.0154 |
| 9 | 0.0043 | 0.0082 |
| 10 | 0.0023 | 0.0043 |
| 11 | 0.0012 | 0.0023 |

| 12 | 0.0006 | 0.0012 |
|----|--------|--------|

## 13 Conclusions

The steps to set up an EMCCD for an observation is as follows:

1) Measure the clock-induced charge $\lambda_{CIC}$ by amplifying dark frames.
2) Use Equation (19) or (20) to find the count rate per frame $\lambda$.
3) Choose a target SNR from science requirements.
4) Use equations (22) and (24) to find the number of exposures $N$ required.
5) Estimate the photon rate $\eta$ (from radiometry).
6) Check if dynamic range (e.g., bright stars, stability) requires adjusting $N$ and thus $\lambda$.
7) Use Equation (15) to find the total observation time $T$.
8) Use Equation (21) to check if standard mode has superior SNR.
9) Adjust $T$ depending on maximum frame rate (PC) and cosmic rate tolerance (SM).
10) Choose amplification gain $G$ based on device operation considerations.
11) Measure the readout noise $\mu$ by reading out unamplified dark frames.
12) Use Equation (41) to find the optimal cutoff $\tau$.
13) Use Table 1 to find the compression code length and Figure 5 for the compression ratio.

Following these steps specifies the operating mode (PC or standard mode), expected SNR, optimal exposure time, and data volume. For a space-based astronomical observatory, knowing these parameters is critical to mission design.

bandwidth compression of visual image signals," *Proceedings of the IEEE* 51(11), 1507-1517 (1963).

**Acknowledgements**

This research was carried out at the Jet Propulsion Laboratory, California Institute of Technology, under a contract with the National Aeronautics and Space Administration (80NM0018D0004). © 2023. California Institute of Technology. Government sponsorship acknowledged.